%% file: sample-authordraft.tex
\documentclass[sigconf]{acmart}
\AtBeginDocument{%
  \providecommand\BibTeX{{%
    \normalfont B\kern-0.5em{\scshape i\kern-0.25em b}\kern-0.8em\TeX}}}

\setcopyright{acmcopyright}
\copyrightyear{2018}
\acmYear{2018}
\acmDOI{10.1145/1122445.1122456}

\acmConference[]{}{}{}
\acmBooktitle{}
\acmPrice{15.00}
\acmISBN{978-1-4503-XXXX-X/18/06}

\begin{document}

\title{Information retrieval for label noise document ranking by bag sampling and group-wise loss}

\author{Chunyu Li}
\email{lcy081099@gmail.com}
\affiliation{%
  \institution{tongji university}
  \country{china}
}

\author{Jiajia Ding}
\email{djjshowtime@gmail.com}
\affiliation{%
  \institution{Edinburgh university}
    \country{china}
}

\author{xing hu}
\email{xinghu1995@outlook.com}
\affiliation{%
  \institution{University College London}
    \country{china}
}
\author{fan wang}
\email{wangfan_1992@126.com}
\affiliation{%
  \institution{tongji university}
    \country{china}
}

\begin{abstract}

Long Document retrieval (DR) has always been a tremendous challenge for reading comprehension and information retrieval. The pre-training model has achieved good results in the retrieval stage and Ranking for long documents in recent years. However, there is still some crucial problem in long document ranking, such as:data label noises, long document representations, negative data Unbalanced sampling, etc. To eliminate the noise of labeled data and to be able to sample the long documents in the search reasonably negatively, we propose the bag sampling method and the group-wise Localized Constrastive Esimation(LCE) method. We use the head middle tail passage for the long document to encode the long document, and in the retrieval, stage Use dense retrieval to generate the candidate's data. The retrieval data is divided into multiple bags at the ranking stage, and negative samples are selected in each bag. After sampling, two losses are combined. The first loss is LCE. To fit bag sampling well, after query and document are encoding, the global features of each group are extracted by convolutional layer and max-pooling to improve the model’s resistance to the impact of labeling noise, finally, calculate the LCE group-wise loss. Notably, our model shows excellent performance on the MS MARCO Long document ranking leaderboard.
\end{abstract}



\maketitle

\section{Introduction}


In recent years, with the development of deep learning, more and more reading comprehension data sets have been released. The more classic ones are SQuAD and SQuAD2.0. Due to the harsh requirements of the industry on data sets, large-scale reading comprehension data sets Start publishing.
Reading comprehension data set MS MARCO (Microsoft Machine Reading Comprehension) based on large-scale real scene data. The data set is based on actual search queries in the Bing search engine and Cortana intelligent assistant and contains 1 million queries, 8 million documents, and 180,000 manually edited answers.


Retrieving and sorting long documents in all data sets is a long document retrieval and sorting task. The search and sorting of long documents are different from the sorting task of short documents, and the difficulty brought by it is as follows. 1) How to encode long documents? Long documents usually contain long words. At the same time, the topic of long documents is difficult to express, which brings challenges to search. 2) Whether it is a task of super long documents or short documents sorting, how to better Reasonable connection of search and sorting. Information retrieval is divided into full ranking and two-stage ranking. The full ranking can effectively avoid the gap caused by two stages.


Through experiments and related theoretical research, it can be concluded that the pipeline using first retrieval + re-ranking can effectively improve the ranking index, but in the actual data set and model, two problems will arise (1): The data set usually has extreme labels -bias For example, through the analysis of the MSMARCO data set and other data sets are shown in fig 1, some of the top-ranked samples can answer the questions accurately, but there is only one marked as a positive sample. Random sampling ng hard will also cause noise Introduced. (2) Experiments have proved that by re-ranking the top data obtained by dense retrieval, good results can be achieved. At the same time, the problem faced is how to effectively inherit the first retrieval ranking information when selecting training samples. The most common method is to normalize the first retrieval score and add query-document as additional features. This paper proposes a method of bag negative sample selection, combined with group-wise loss calculation. On the one hand, it can alleviate the positive and negative sampling boundary problem caused by label bias. At the same time, the group-wise loss formed by the bag can inherit the sorting of first retrieval information.

\section{Related Work}

\subsection{Full Ranking And Re-ranking Model}
In the field of information retrieval, there have been many machine learning ranking models (Learning to Rank) used to solve document ranking problems in the early days, including LambdaRank\cite{burges2010ranknet}, AdaRank\cite{xu2007adarank}, etc. These models rely on many manually constructed features. With the popularity of deep learning technology in machine learning, researchers have proposed many neural ranking models, such as DSSM \cite{dssmshen2014learning}, KNRM \cite{KNRMdai2018convolutional}, and so on. These models map the representation of the problem and the document into a continuous vector space and then calculate their similarity through a neural network, thus avoiding tedious manual feature construction.

In recent years, pre-trained models have successfully brought significant progress to the NLP field. An important goal of pre-training models such as GPT\cite{GPTradford2018improving}, XLNET\cite{xlnetyang2019xlnet}, and BERT\cite{BERTdevlin2018bert} is to pre-train through a large amount of unlabeled data. Through large-scale, The label data and the training model can be greatly improved by only a small number of labeled data sets in the subsequent finetuning stage.
BERT is the first to start the pre-training model. It uses a two-way transformer coding structure to perform MLM and NSP tasks to obtain contextual language representations and sentence-pair representations. Applying BERT directly to downstream IR tasks can achieve better results. In addition to now using BERT for IR tasks, there have also been many pre-training models specifically for IR tasks.

In order to coordinate efficiency and contextualization in IR, the author proposes a ColBERT ranking model based on contextualized post-interaction. ColBERT proposed a new post-interaction paradigm for estimating the correlation between query q and document d. Compared with the existing bert-based model, the results show that ColBERT is more than 170 times faster and requires FLOPs /query is 14000 times less, and has little impact on quality, and is better than all non-bert baselines.

BM25 has always been a strong baseline. However, for semantic recall and sorting, the accuracy of BM25 is greatly reduced. DPR\cite{dprkarpukhin2020dense} is a search framework based on vector representation. The article proves that retrieval based on dense vectors is superior to the traditional sparse retrieval in answering open-domain questions. Experiments show that a simple dual encoder method can achieve amazing results. More complex model frameworks or similarity measurement functions may not provide additional gains. DPR improves the results of retrieval performance and breaks records in multiple open domain question answering benchmark tests

ANCE proposes Approximate nearest neighbor Negative Contrastive Learning\cite{ancexiong2020approximate}, a learning mechanism that selects hard training negatives globally from the entire corpus, using an asynchronously updated ANN index. Demonstrate the effectiveness of ANCE on web search, question answering, and in a commercial search environment.

\subsection{Long Document Representation And Ranking}
Although BERT is present, due to the design of the BERT model itself, it faces some challenges when processing texts longer than 512 tokens. Therefore, it is difficult to apply BERT to the document ranking task directly. Specifically, it is mainly reflected in the following two aspects 
1)During the training phase, we still don't know what form of content to provide to the model. The critical issue is that the data set at the document level provides relevant judgments for document ranking. However, the assessment of "relevance" comes from the document that contains "relevant content", but we cannot accurately judge whether the correct answer is the subject of the document and how it is distributed in the document. 
2) In the inference phase, if the document is too long to be fully inputted to BERT, how to deal with the long document becomes the primary problem. At the same time, in the training phase and the inference phase, different processing methods will produce different gaps. 

Birch\cite{brichyilmaz2019cross} used data without length problems for training and then migrated these correlation matching models to the target domain/task, thereby completely avoiding training problems
In the prediction phase, estimating the relevance of a document is converted to evaluating the relevance of a single sentence, and then the result scores are aggregated. Birch needs to calculate the document relevance, inferring each sentence in the document, and then top-n Combination of the score and original document score (i.e., first stage score)
Among them is the score based on the highest scoring sentence of BERT, and the inference for each sentence is the same. In other words, the relevance score of the document comes from the original candidate document score (for example, from BM25) and the group of the highest-scoring sentence in the document determined by the BERT model.

As an extension of BERT–MaxP\cite{MAXPdai2019deeper}, PCGM\cite{PCGMwu2020leveraging} considers whether grading the paragraph-level relevance judgments during training can get a more effective ranking. In response to this problem, the author marked the paragraph-level cumulative income in the Chinese news corpus, which is defined as the amount of relevant information that readers will accumulate after reading the document to the specified paragraph. This work uses natural paragraphs as paragraphs. Therefore, by definition, the cumulative gain at the document level is the incremental gain at the highest paragraph level. Based on these human annotations, the author found the following conclusion: On average, highly relevant documents are longer than other types of documents, whether in paragraphs or words. The higher the cumulative gain at the document level, the more paragraphs the user needs to read before reaching the cumulative growth at the document level through the incremental increase at the paragraph level 

Document-level sorting based on paragraph representation aggregation PARADE\cite{PARADEli2020parade} is a series of models that can divide the long text into multiple paragraphs and summarize each paragraph's [CLS] representation. Specifically, PARADE divides the long text into a fixed number of fixed-length paragraphs. When the text contains fewer paragraphs, the paragraphs will be filled and blocked during the presentation aggregation process. When the text contains more paragraphs, the first and last paragraphs are always kept, and the remaining paragraphs will be randomly sampled. Successive paragraphs partially overlap to minimize the chance of separating relevant information from its context.

\subsection{Negative Sampling and Loss Function For Ranking Model}
The data space for a retrieval task has various data distribution in degrees of text/semantic/social matches, and it is important to design a training data set for an embedding model to learn efficiently and effectively on such space\cite{huang2020embedding}. Much research effort has addressed sampling strategies for hard negative and hard positive. Positive
sampling techniques have also been proposed for sentence, audio\cite{logeswaran2018efficient}\cite{oord2018representation}. In supervised learning, the hard negative sample can help guide a model to correct mistakes more quickly\cite{oh2016deep}, \cite{robinson2020contrastive} introduce a simple hard negative sampling method overcoming an apparent roadblock: negative mining in metric learning uses pair-wise similarity information as a core component, while contrastive learning is unsupervised. 
The loss function is an important module in the Ranking model, and There is always a three-loss function in the Ranking model point-wise, pair-wise and list-wise. Point-wise sorting takes each item in the training set as a sample to obtain the rank function. The main solution is to convert the classification problem into a classification or regression problem for a single item. Pair-wise sorting uses two different items in the same query as a sample. The main idea is to convert the rank problem into a binary classification problem. Listwise sorting regards the entire item sequence as a sample and is realized by directly optimizing the evaluation method of information retrieval and defining the loss function. These loss functions need according to different applications to choose.\cite{liu2011learning}

\section{THE PROPOSED MODEL}
In this section, in response to the above problems, bag-sampling and group-wise loss for long document ranking are proposed. The structure of the model is shown in the figure. First, it introduces how to model the long document, and secondly, for the selection of negative samples, use a bag. The method, under the condition of dividing the difficulty level, minimizes the noise caused by fewer labeled samples. Finally, to be able to inherit the scores provided by the first retrieval and maximize the effect of bag sampling, use LCE group-wise loss to calculate the loss score of each step group. The model architecture, as shown in Fig 1 and Fig 2
\begin{figure}[h]
  \centering
  \includegraphics[width=\linewidth]{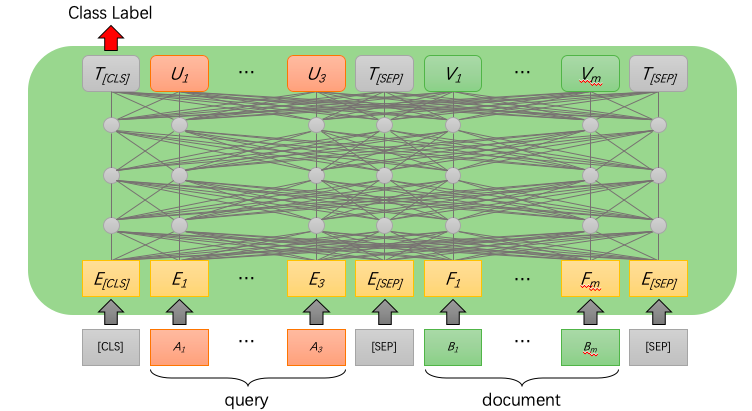}
  \caption{bert question answer}
  \Description{bag sampling analysis}
\end{figure}
\begin{figure}[h]
  \centering
  \includegraphics[width=\linewidth]{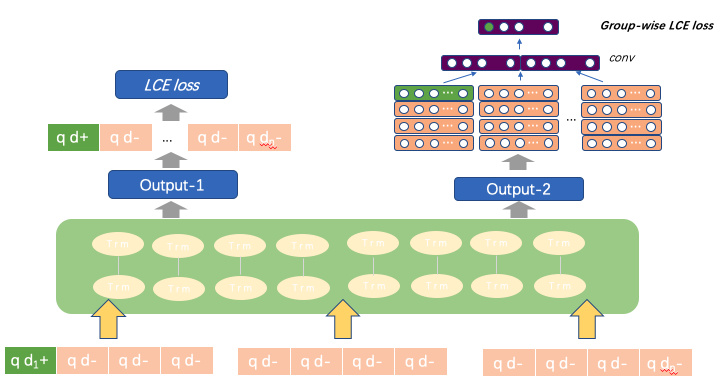}
  \caption{model architecture}
  \Description{bag sampling analysis}
\end{figure}

\subsection{Long Document Model}
This paper designs the following long text expressions based on the statistical distribution of long text lengths and related answer distributions. The maximum size of bert is 512. Through analysis, it can be seen that some answers exist at the end and the center of the sentence. Usually, the central theme of long documents is the title, the beginning of the article, and the conclusion of the end. Inspired by maxp, this article will design the head-tail passage coding, maxp is used to divide the long document into four parts, this article divides the document into two necessary parts of the head and tail, and the remaining two Passage Randomly select two passages of 512 lengths in the center, shown as in fig 2. The final results in equation(1) (2)
\begin{equation}
    Score_{document} = [S_{head},S_{middle-1},S_{middle-2},S_{tail}]
\end{equation}

\begin{equation}
    Score(passage_{i}) = cls(BERT(concat(q, passage_{i})))
\end{equation}

\subsection{BAG Sampling for Reranking}

Typically, when deep learning is sampling, it uses the same batch of data for the sample and calculates the gradient. That is, for the same query q, the recall strategy is used to randomly select the document d+ related to q from the recalled document $D^retrieval$, and more Unrelated documents d-, as shown in equation.

\begin{equation}
\begin{split}
        Batch = {(q, d^+, d^-)|q\in rand(Q^N), 
        d^+, d^-\in rand(D^{retrieval})}
\end{split}
\end{equation}

ANCE uses global negative sampling to dynamically sample negative samples and sample some hard negative samples simultaneously. It is challenging to select negative samples, and for incompletely labeled data sets (such as MSMARCO), the correct labeled answer for each question is only One. We selected a question as an example as shown in Fig 3
From the clustering graph, it can be concluded that the query document after the bert encoder will show apparent category effects. Since there is only one correctly labeled answer, there will be documents and topics similar to the correct answer near the right answer. It is more difficult to distinguish, and in the case of a small labeled data set, the documents with similar valid documents are likely to be accurate but unlabeled data. This article will carry out bag sampling for the interval section, which can effectively distinguish the samples at the stage The set is N ($d_1 ,d_2,d_3,d_4,d_5.....d_N$ ), the sample set is divided into M bags, (1, 2, 3, 4, 5....M), randomly select i samples in each pack for calculation:
\begin{equation}
        Batch_{bs} = {(q, B(d^+_{1}, d^-_{5}),B(d^-_{6}, d^-_{8})..B(d^-_{N-1}, d^-_{N}))}
\end{equation}

\begin{equation}
        Batch_{bs} = {q\in rand(Q^N), B\in M, d^+, d^-\in rand(B_N)}
\end{equation}

\begin{figure}[h]
  \centering
  \includegraphics[width=\linewidth]{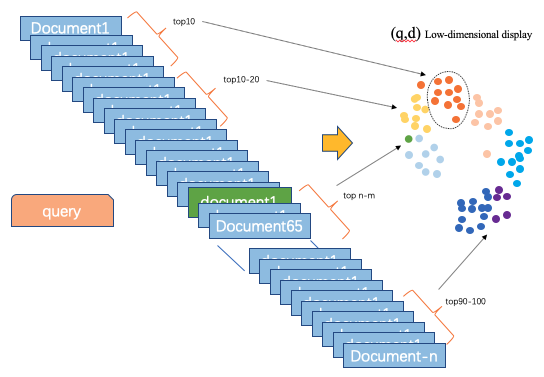}
  \caption{bag sampling analysis}
  \Description{bag sampling analysis}
\end{figure}
\subsection{LCE for Group Wise Loss}
\subsubsection{{\textbf Local Contrastive Loss}}
Inspired by LCE, this article uses the first stage recall topN data set, uses the ANCE method, and uses local negative sampling to calculate the correlation between a single accurate sample and the document. The specific sampling method is as shown in section 3.2 to select the bag sampling data. 
 After selecting all positive and negative samples, extract the group of positive samples to the top, and calculate contractual loss, as shown in equation 5

\begin{equation}
        Score(q,d) = V^T_{p}CLS(BERT(concat(q, d)))
\end{equation}
with which we define the contrastive loss for each query q as:
\begin{equation}
        \boldsymbol{L_{q,d}} = -\log \frac{exp(score(q,d^+_q))}{\sum_{d\in G_q} exp(score(q,d^-_q))}
\end{equation}

\subsubsection{Local Contrastive Group Wise Loss}


To make bag negative sampling more useful in reranking, this article proposes LCGW(local contraste group wise ) loss for the first time. After encoding by bert, the expression of each group is obtained according to Fig 2. $[(r(q,d+),r(q, d-),r(q, d-)),(r(q,d-),r(q,d-),r(q,d-))]$, To obtain the global characteristics of each group, we use a layer of convolutional network to extract the overall signal characteristics of each group, then max-pool selects the strong high level matching signal characteristics, and finally gets Score of each group. Through group high-level feature extraction, we can get that the feature representations in each group are similar. This kind of global feature representation helps to remove the influence caused by insufficient labeling.

\begin{equation}
        Input_{Group-i}= concat(BERT(q,d_1)...BERT(q,d_i))
\end{equation}

\begin{equation}
        \boldsymbol{Encoder_{Group-i}} = Conv(Input_{Group-i})
\end{equation}


%


\section{EXPERIMENTS}
Our main training and inference dependency is PyTorch,
HuggingFace transformer,in the initial recall stage, we use ANCE to recall the candidate set
\subsection{Document collection and query Datasets}
There is insufficient labeling, and large-scale, long document data sets are relatively scarce. MSMRACO is a pretty typical data set, with 300W long documents and 30w training annotation queries. Due to the massive amount of data, it is easy to appear in the literature above. The labeling is not sufficient, and labeling noise is generated. As an indicator, we select the commonly used MRR as the evaluation indicator of the test data.
In the initial stage of training, the 30w query data is trained and recalled through ANCE, and the number of recalls is top100, and the sampling method proposed in this article is used for sampling.
\subsection{Parameter For Model and Results}
We use topk k=100 retrieval document for Re-ranking, group size 10 and each bag size s 10, sample 4 documents from bag, the  results as shown in Table 1, our model achieves the best results on the dev dataset, and it drops slightly on the test
\begin{table}
  \caption{RESULTS ANCE  BAG SAMPLING GROUP-WISE LOSS(ANCE BS+GL) LEADER BOARDER MSMARCO-Document-Ranking 2021/04/27}
  \label{tab:freq}
  \begin{tabular}{llll}
    \toprule
    Model & MRR@100(dev) & MRR@100(eval) \\
    \midrule
    DML & 0.470 & 0.416\\
    LCEB Model & 0.479 & 0.419 \\
    PROP step400K(v2) & 0.479 &  \bfseries{0.423}\\
    ANCE + LongP (ensemble) & 0.481 & 0.420 \\
    \bfseries{ANCE BS+GL} & \bfseries{0.489} & 0.421\\
  \bottomrule
\end{tabular}
\end{table}
\section{CONCLUSION}
This paper proposes the Long document ranking model, which use max-passage and bag sampling method, to avoid hard negative and positive affect, the model adopt group-wise loss . Of course, there are still some shortcomings. We will test the model's effect on more datasets in the future and explore the influence of more sampling mechanisms. 

\section{Acknowledgments}

\input{sample-authordraft.bbl}

\bibliographystyle{ACM-Reference-Format}
\bibliography{sample-authordraft}

\end{document}